\documentstyle[aps]{revtex}
\tightenlines
\begin{document}

\title{Complete High Temperature Expansions
for One-Loop Finite Temperature Effects}

\author{Peter N. Meisinger and Michael C. Ogilvie}
\address{ Dept. of Physics
Washington University
St. Louis, MO 63130}

\date{\today}
\maketitle

\begin{abstract}

We develop exact, simple closed form expressions for
partition functions associated with relativistic bosons and fermions in odd
spatial dimensions. These expressions, valid at high temperature, include
the effects of a non-trivial Polyakov loop and generalize well-known high
temperature expansions. The key \ technical point is the proof of a set of
Bessel function identities which resum low temperature expansions into high
temperature expansions. The complete expressions for these partition
functions can be used to obtain one-loop finite temperature contributions to
effective potentials, and thus free energies and pressures.

\end{abstract}

\widetext

\section{\protect\vspace{1pt}Introduction}

The applications of finite temperature field theory are numerous and
diverse \cite{Kapusta:1989tk,Bellac:1996,Das:1997gg}  
For many applications, a high-temperature expansion of one-loop
contributions to thermodynamic functions is necessary. A typical one-loop
contribution to a $(d+1)$-dimensional effective potential from a bosonic
degree of freedom and its antiparticle has the form 
\begin{equation}
V=\frac{2}{\beta }\int \frac{d^{d}k}{\left( 2\pi \right) ^{d}}\ln \left[
1-e^{-\beta \omega _{k}}\right] 
\end{equation}
where the relativistic energy $\omega _{k}$ is given by $\sqrt{k^{2}+M^{2}}$%
. In many cases, the mass $M$ is a function of other quantities, most
notably vacuum expectation values of fields. When the total effective
potential attains its minimum, $V\,$\ may be identified as a contribution to
the total free energy, and $-V$ as a contribution to the pressure. A more
general case is obtained when there is a non-trivial, but spatially uniform,
Polyakov loop as well as a non-zero chemical potential $\mu $.In this case
we have 
\begin{equation}
V_{B}\left( \theta -i\beta \mu \right) =\frac{1}{\beta }\int \frac{d^{d}k}{%
\left( 2\pi \right) ^{d}}\ln \left[ 1-e^{-\beta \omega _{k}+i\theta +\beta
\mu }\right] +\frac{1}{\beta }\int \frac{d^{d}k}{\left( 2\pi \right) ^{d}}%
\ln \left[ 1-e^{-\beta \omega _{k}-i\theta -\beta \mu }\right] \text{.} 
\end{equation}
Note that the effect of a non-trivial Polyakov loop is to add a phase factor $%
\exp \left( \pm i\theta \right) $ to $\exp \left( -\beta \omega \right) $.\
We will evaluate $V_{B}$ for arbitrary $\theta $. In principle, the effect
of $\mu $ can be included by a careful analytic continuation $\theta
\rightarrow \theta -i\beta \mu $, but we do not consider it here. 

We
will develop a high temperature expansion for $V_{B}\left( \theta \right) $,
valid for $d\,$odd, as well as a similar expression for the corresponding
fermionic quantity $V_{F}(\theta )$. This derivation is simple and exact,
and generalizes the results of Dolan and Jackiw \cite{Dolan:1974qd},
who gave approximate high-temperature
expressions for $V_{B}(\theta=0 )$ and $V_{F}(\theta=0 )$
valid up to order $M^4 ln(\beta M)$ in four dimensions.
The work of Dolan and Jackiw was extended by Haber and Weldon
\cite{Haber1:1982} \cite{Haber2:1982}
who gave a complete expression for the Bosonic case $V_B$ as an infinite sum
over hypergeometric functions. Their work included a non-zero chemical
potential. Later work by Actor showed that similar high-temperature expansions
could be obtained using zeta-function techniques
\cite{Actor:1986zf,Actor:1987cf}.
In both cases, higher-order correction terms are given by infinite series
in $\beta M$. 
Our expressions effectively resum these corrections into
a simpler form. 
Analytical results for 
the case of a non-trivial Polyakov loop, $\theta \neq 0$, were first given
in the case $M=0$ by Gross, Pisarski and Yaffe
and by Weiss \cite{Gross:1981br,Weiss:1981rj,Weiss:1982ev}.
Our work generalizes their results to the case $M \neq 0$.
The higher-order terms in
our expressions are manifestly periodic
in $\theta$.
This periodicity
is important in the application
of these results to our recent work with Miller on
models of the deconfinement transition
\cite{Meisinger:2001cq}.
In this work, the eigenvalues of the Polyakov loop
serve as the order parameters for deconfinement,
a point of view also emphasized recently by
Pisarski\cite{Pisarski:2000eq}.

Before beginning the
derivation, we give some examples of its application.
As a first example, consider a scalar boson in the fundamental
representation of an $SU(N)$ gauge group. The Polyakov loop is an $N\times N$
unitary matrix given in general by 
\begin{equation}
P\left( \overrightarrow{x}\right) ={\cal T} \exp 
\left[ i\int_{0}^{\beta
}d\tau \,A_{0}\left( \overrightarrow{x},\tau \right) \right] 
\end{equation}
where ${\cal T}$ on the right-hand 
side indicates Euclidean time ordering. Here we
assume that the Polyakov loop can be made spatially uniform by an
appropriate choice of gauge. A global unitary transformation then puts $P$
into the diagonal form 
\begin{equation}
P_{jk}=\delta _{jk}\exp \left( i\theta _{j}\right) 
\end{equation}
and the partition function is 
\begin{equation}
\sum_{j}V_{B}\left( \theta _{j}\right) \text{.} 
\end{equation}

As a second example,
consider the case of the gauge bosons themselves, which lie in the
adjoint representation of the gauge group. The Polyakov loop in the adjoint
representation is an $\left( N^{2}-1\right) \times \left( N^{2}-1\right) $
matrix. The partition function for the $N^{2}-1$ particles is 
\begin{equation}
s\frac{1}{2}\sum_{j,k=1}^{N}(1-\frac{1}{N}\delta _{jk})V_{B}\left( \theta
_{j}-\theta _{k}\right) 
\end{equation}
where the $\delta _{jk}$ removes a singlet contribution, and the factor of $%
1/2$ corrects for overcounting since $V_{B}$ has both a particle and
antiparticle contribution. The factor $s$ accounts for spin degeneracy; in $%
3+1$ dimensions $s=2$, a consequence of the two possible polarization states
of gauge bosons .

For our third and final example, consider
the evaluation of fermionic partition functions, which can be reduced to the
general bosonic problem. A typical fermionic contribution of particle and
antiparticle has the form 
\begin{equation}
V_{F}\left( \theta -i\beta \mu \right) =-\frac{1}{\beta }\int \frac{d^{d}k}{%
\left( 2\pi \right) ^{d}}\ln \left[ 1+e^{-\beta \omega _{k}+i\theta +\beta
\mu }\right] -\frac{1}{\beta }\int \frac{d^{d}k}{\left( 2\pi \right) ^{d}}%
\ln \left[ 1+e^{-\beta \omega _{k}-i\theta -\beta \mu }\right] 
\end{equation}
which is easily written as 
\begin{equation}
V_{F}\left( \theta \right) =-V_{B}\left( \pi +\theta \right) \text{.} 
\end{equation}
For fermions in the fundamental representation of $SU(N)$, the partition
function is 
\begin{equation}
s\sum_{j}V_{F}\left( \theta _{j}\right) =-s\sum_{j}V_{B}\left( \pi +\theta
_{j}\right) 
\end{equation}
where the factor $s$ again accounts for spin degeneracy.

In section 2, we review the derivation of low temperature expansions for $%
V_{B}\left( \theta \right) $ and $V_{F}\left( \theta \right) $. Section 3
derives the Bessel function identities which convert these low temperature
expansions to high temperature expansions. Section 4 applies these
identities to the case of three spatial dimensions. A final section gives
brief conclusions. There are two appendices.

\vspace{1pt}

\section{Low Temperature Expansion in $d$ Dimensions}

\vspace{1pt}A low-temperature expansion for $V_{B}\left( \theta \right) $
can be generated for arbitrary spatial dimension $d$ by expanding the
logarithm and integrating term by term, first over the surface of a $d$%
-dimensional sphere, and then over a radial degree of freedom $k$ 
\cite{Actor:1986zf,Actor:1987cf}.
The result, given in terms of modified Bessel functions
\cite{GandR}, is
\begin{eqnarray}
V_{B}\left( \theta \right) &=&\frac{1}{\beta }\int \frac{d^{d}k}{\left( 2\pi
\right) ^{d}}\ln \left[ 1-e^{-\beta \omega _{k}+i\theta }\right] +\frac{1}{%
\beta }\int \frac{d^{d}k}{\left( 2\pi \right) ^{d}}\ln \left[ 1-e^{-\beta
\omega _{k}-i\theta }\right] \\
&=&-\frac{M^{d/2+1/2}}{2^{d/2-3/2}\pi ^{d/2+1/2}\beta ^{d/2+1/2}}%
\sum_{n=1}^{\infty }\frac{1}{n^{d/2+1/2}}K_{\left( d+1\right) /2}\left(
n\beta M\right) \cos \left( n\theta \right)
\end{eqnarray}
which is derived in detail in Appendix A. Each term in the series represents
the contribution of $n$ particles or antiparticles, with a corresponding
phase\ factor of $\exp \left( \pm in\theta \right) $. If the one-loop finite
temperature functional determinant is represented as a functional integral
over a space-time variable $x_{\mu }$, the phase factors are associated with
paths which wind non-trivially in the Euclidean time direction.

For fermions, we have 
\begin{equation}
V_{F}\left( \theta \right) =\frac{M^{d/2+1/2}}{2^{d/2-3/2}\pi
^{d/2+1/2}\beta ^{d/2+1/2}}\sum_{n=1}^{\infty }\frac{\left( -1\right) ^{n}}{%
n^{d/2+1/2}}K_{\left( d+1\right) /2}\left( n\beta M\right) \cos \left(
n\theta \right) \text{.} 
\end{equation}
In a path integral representation, the factors of $\left( -1\right) ^{n}$
are a consequence of fermionic antiperiodic boundary conditions. We next
derive a set of identities which resum these series for $d$ odd.

\section{\protect\vspace{1pt}Bessel Function Identities}

We will derive a set of identities for sums of the form 
\begin{equation}
\sum_{n=1}^{\infty }\frac{1}{n^{p}}K_{p}\left( nr\right) \cos \left( n\phi
\right) 
\end{equation}
\vspace{1pt}for $p$ an even integer. Our starting point is the identity 
\begin{equation}
\sum_{p=1}^{\infty }K_{0}(pr)\cos (p\phi )=\frac{1}{2}\left[ \gamma +\ln
\left( \frac{r}{4\pi }\right) \right] +\frac{\pi }{2}\sum_{l}\,^{^{^{\prime
}}}\left[ \frac{1}{\sqrt{r^{2}+\left( \phi -2\pi l\right) ^{2}}}-\frac{1}{%
2\pi \left| l\right| }\right] 
\end{equation}
which may be found in \cite{GandR};
we provide a derivation in Appendix B which
provides some physical insight into its origin. The notation $%
\,\sum_{l}\hspace{0.00in}^{^{^{\prime }}}\,$is 
used to indicate that singular terms, here
the $1/\left| l\right| $ term, are omitted when $l=0$.

Using the recursion formula 
\begin{equation}
\frac{d}{dz}K_{\nu }(z)=-K_{\nu -1}(z)-\frac{\nu }{z}K_{\nu }(z)\text{,} 
\end{equation}
it follows immediately that 
\begin{equation}
\frac{d}{dz}\sum_{p=1}^{\infty }\frac{z}{p}K_{1}(pz)\cos (p\phi
)=-z\sum_{p=1}^{\infty }K_{0}(pz)\cos (p\phi )\text{.} 
\end{equation}
This in turn implies that 
\begin{equation}
\sum_{p=1}^{\infty }\frac{1}{p}K_{1}(pz)\cos (p\phi )=-\frac{1}{z}\int
dz\,\,\,z\left[ \sum_{p=1}^{\infty }K_{0}(pz)\cos (p\phi )\right] +\frac{%
C(\phi )}{z} 
\end{equation}
where $C(\phi )\,$is an unknown function to be determined. Integration
yields immediately 
\begin{equation}
\sum_{p=1}^{\infty }\frac{1}{p}K_{1}(pz)\cos (p\phi )=-\frac{1}{4}z\left[
\ln \left( \frac{z}{4\pi }\right) +\gamma -\frac{1}{2}\right] -\frac{\pi }{2z%
}\sum_{l}\,^{^{^{\prime }}}\left[ \sqrt{z^{2}+\left( \phi -2\pi l\right) ^{2}%
}-\frac{z^{2}}{4\pi \left| l\right| }\right] +\frac{C(\phi )}{z}\text{.} 
\end{equation}
The function $C(\phi )\,$is determined from the behavior of $K_{\nu}(z)$ for $%
z\rightarrow 0$%
\begin{equation}
K_{\nu }(z) \rightarrow  \frac{1}{2}\Gamma (\nu )\left( \frac{2}{z}\right) ^{\nu
} 
\end{equation}
in combination with the standard result \cite{GandR}
\begin{equation}
\label{eq:ber2}
\sum_{p=1}^{\infty }\frac{\cos (p\phi )}{p^{2}}=\frac{1}{4}\phi ^{2}-\frac{%
\pi }{2}\phi +\frac{\pi ^{2}}{6}\text{,} 
\end{equation}
which is valid for $0\leq \phi <2\pi $. The right hand side of 
Eq.\ (\ref{eq:ber2}) is a
rescaling of the second Bernoulli polynomial; it can be extended to all real
values if $\phi \,$is replaced by $\phi\, mod\, 2\pi $ on the right hand
side of the equation. This implies the leading behavior of the sum as $%
z\rightarrow 0\,$is given by 
\begin{equation}
\lim_{z\rightarrow 0}\,z\sum_{p=1}^{\infty }\frac{1}{p}K_{1}(pz)\cos (p\phi
)=\sum_{p=1}^{\infty }\frac{\cos (p\phi )}{p^{2}}=\frac{1}{z}\left[ \frac{1}{%
4}\phi ^{2}-\frac{\pi }{2}\phi +\frac{\pi ^{2}}{6}\right] \text{,} 
\end{equation}
giving us finally 
\begin{eqnarray}
\sum_{p=1}^{\infty }\frac{1}{p}K_{1}(pz)\cos (p\phi ) &=&-\frac{1}{4}z\left[
\ln \left( \frac{z}{4\pi }\right) +\gamma -\frac{1}{2}\right] +\frac{1}{z}%
\left[ \frac{1}{4}\phi _{+}^{2}-\frac{\pi }{2}\phi _{+}+\frac{\pi ^{2}}{6}%
\right] \nonumber\\
&&-\frac{\pi }{2z}\sum_{l}\,^{^{^{\prime }}}\left[ \sqrt{z^{2}+\left( \phi
-2\pi l\right) ^{2}}-\left| \phi -2\pi l\right| -\frac{z^{2}}{4\pi \left|
l\right| }\right]
\end{eqnarray}
where we have introduced the notation $\phi _{+}$ to represent $\phi \,
mod\, 2\pi $. When discussing fermions, we will also use $\phi _{-}$ to
similarly represent an angle chosen to lie between $-\pi $ and $\pi $. Note
that the last part of this expression is automatically periodic due to the
sum over $l$.

Application of this technique a second time gives 
\begin{eqnarray}
\sum_{p=1}^{\infty }\frac{1}{p^{2}}K_{2}(pz)\cos (p\phi ) &=&\frac{1}{16}%
z^{2}\left[ \ln \left( \frac{z}{4\pi }\right) +\gamma -\frac{3}{4}\right] -%
\frac{1}{2}\left[ \frac{1}{4}\phi _{+}^{2}-\frac{\pi }{2}\phi _{+}+\frac{\pi
^{2}}{6}\right]  \nonumber\\
&&+\frac{2}{z^{2}}\left[ \frac{-1}{48}\phi _{+}^{4}+\frac{\pi }{12}\phi
_{+}^{3}-\frac{\pi ^{2}}{12}\phi _{+}^{2}+\frac{\pi ^{4}}{90}\right]\nonumber\\
&&+\frac{\pi }{2z^{2}}\sum_{l}\,^{^{^{\prime }}}\left\{ \frac{1}{3}\left[
z^{2}+\left( \phi -2\pi l\right) ^{2}\right] ^{3/2}-\frac{1}{3}\left| \phi
-2\pi l\right| ^{3}-\frac{1}{2}\left| \phi -2l\pi \right| z^{2}-\frac{z^{4}}{%
16\pi \left| l\right| }\right\} 
\end{eqnarray}
which is needed for the case $d=3$. We have used the standard result 
\cite{GandR}
\begin{equation}
\sum_{p=1}^{\infty }\frac{\cos (p\phi )}{p^{4}}=\frac{-1}{48}\phi _{+}^{4}+%
\frac{\pi }{12}\phi _{+}^{3}-\frac{\pi ^{2}}{12}\phi _{+}^{2}+\frac{\pi ^{4}%
}{90}\text{.}
\end{equation}
Formulas appropriate for $d=5,7,..$ can also be derived in the same manner.

\section{\protect\vspace{1pt}High-Temperature Expansions for $d=3$}

\vspace{1pt}We now can write complete expressions for $V_{B}\left( \theta
\right) $ and $V_{F}\left( \theta \right) $ in three spatial dimensions:
\begin{eqnarray}
V_{B}\left( \theta \right)  &=&-\frac{M^{2}}{\pi ^{2}\beta ^{2}}%
\sum_{n=1}^{\infty }\frac{1}{n^{2}}K_{2}\left( n\beta M\right) \cos \left(
n\theta \right) \nonumber \\
&=&-\frac{2}{\pi ^{2}\beta ^{4}}\left[ \frac{\pi ^{4}}{90}-\frac{1}{48}%
\theta _{+}^{4}+\frac{\pi }{12}\theta _{+}^{3}-\frac{\pi ^{2}}{12}\theta
_{+}^{2}\right] +\frac{M^{2}}{2\pi ^{2}\beta ^{2}}\left[ \frac{1}{4}\theta
_{+}^{2}-\frac{\pi }{2}\theta _{+}+\frac{\pi ^{2}}{6}\right] \nonumber \\
&&-\frac{1}{2\pi \beta ^{4}}\sum_{l}\,^{^{^{\prime }}}\left\{ \frac{1}{3}%
\left[ \left( \beta M\right) ^{2}+\left( \theta -2\pi l\right) ^{2}\right]
^{3/2}-\frac{1}{3}\left| \theta -2\pi l\right| ^{3}-\frac{1}{2}\left| \theta
-2l\pi \right| \beta ^{2}M^{2}-\frac{\left( \beta M\right) ^{4}}{16\pi
\left| l\right| }\right\} \nonumber \\
&&-\frac{M^{4}}{16\pi ^{2}}\left[ \ln \left( \frac{\beta M}{4\pi }\right)
+\gamma -\frac{3}{4}\right] \text{.}
\end{eqnarray}
Parts of this complete expression have been known for some time. For $\theta
=0$, the leading behavior is 
\begin{equation}
V_{B}\left( \theta =0\right) \approx -2\frac{\pi ^{2}}{90\beta ^{4}}+\frac{%
M^{2}}{12\beta ^{2}}-\frac{M^{3}}{6\pi \beta }-\frac{M^{4}}{16\pi ^{2}}%
\left[ \ln \left( \frac{\beta M}{4\pi }\right) +\gamma -\frac{3}{4}\right] 
\text{,}
\end{equation}
an expression first derived by Dolan and Jackiw \cite{Dolan:1974qd}.
The leading $T^{4}$
behavior for $\theta \neq 0$ was first derived by Gross, Pisarski and Yaffe
and Weiss for $SU(N)$ gauge bosons and for massless fermions
\cite{Gross:1981br,Weiss:1981rj,Weiss:1982ev}. 

Each of the four terms deserves comment. The first is the blackbody free
energy for two degrees of freedom, and depends only on the temperature and
the angle $\theta $. The second term, which is the leading correction due to
the mass $M$, often appears in discussions of symmetry restoration at high
temperatures with $\theta =0$. For example, suppose we are calculating the
effective potential for a complex scalar field $\Phi $. The mass $M^{2}$ is
given by the second derivative of the classical potential, $\partial
^{2}U/\partial \Phi ^{\ast }\partial \Phi $, and depends on the expectation
value of the field $\Phi $. If $U$ has the form $U=-m^{2}\Phi ^{\ast }\Phi
+\lambda \left( \Phi ^{\ast }\Phi \right) ^{2}$, then $M^{2}=-m^{2}+4\lambda
\Phi ^{\ast }\Phi $. For $m^{2}>0$, the $U(1)$ symmetry is spontaneously
broken at low temperature. At high temperature, the $M^{2}/12\beta ^{2}$
term generates a positive mass for the $\Phi $ field of order $T$, restoring
the symmetry. The third term in $V_{B}\left( \theta \right) $ is linear in $T
$, and non-analytic in $M^{2}$ for $\theta =0$. It is closely associated
with the $n=0$ Matsubara mode, which is the most infrared singular
contribution to a finite temperature functional determinant. This term is
responsible for non-analytic behavior in finite temperature perturbation
theory via the summation of ring diagrams. For example, in a scalar theory
it gives rise to the $\lambda ^{3/2}$ contribution to the free energy; in
QED, the contribution is $e^{3}$ \cite{Kapusta:1989tk}
Note how subtractions occur
in the $l\neq 0$ parts of this term to keep these parts subleading.\ The
last term is logarithmic in the dimensionless combination $\beta M$ and
independent of $\theta $. In calculations of effective potentials, it
typically combines with zero-temperature logarithms in such a way that the
temperature $T$ sets the scale of running coupling constants at high $T$.

From the basic result for $V_{B}$, we can build other results. Consider a
complex scalar field in the fundamental representation of $SU(N)$. The
partition function in a constant background Polyakov loop is given by 
\begin{eqnarray}
V_{FT} &=&\sum_{j}V_{B}\left( \theta _{j}\right) \nonumber\\
&=&-\frac{\pi ^{2}N}{45\beta ^{4}}+\frac{2}{\pi ^{2}\beta ^{4}}%
\sum_{j}\left[ \frac{1}{48}\theta _{j}^{4}-\frac{\pi }{12}\theta _{j}^{3}+%
\frac{\pi ^{2}}{12}\theta _{j}^{2}\right] \nonumber \\
&&+\frac{NM^{2}}{12\beta ^{2}}+\frac{M^{2}}{2\pi ^{2}\beta ^{2}}%
\sum_{j}\left[ \frac{1}{4}\theta _{j}^{2}-\frac{\pi }{2}\theta _{j}\right] 
\nonumber\\
&&-\frac{1}{2\pi \beta ^{4}}\sum_{j,l}\,^{^{^{\prime }}}\left\{ \frac{1%
}{3}\left[ \left( \beta M\right) ^{2}+\left( \theta _{j}-2\pi l\right)
^{2}\right] ^{3/2}-\frac{1}{3}\left| \theta _{j}-2\pi l\right| ^{3}-\frac{1}{%
2}\left| \theta _{j}-2l\pi \right| \beta ^{2}M^{2}-\frac{\left( \beta
M\right) ^{4}}{16\pi \left| l\right| }\right\} \nonumber \\
&&-\frac{NM^{4}}{16\pi ^{2}}\left[ \ln \left( \frac{\beta M}{4\pi }\right)
+\gamma -\frac{3}{4}\right] 
\end{eqnarray}
where we assume all $\theta $'s are chosen to lie between $0$ and $2\pi $ in
accordance with the convention for $\theta _{+}$. A similar expression holds
for bosons in the adjoint representation.

\vspace{1pt}For fermions, we have similarly 
\begin{eqnarray}
V_{F}\left( \theta \right) &=&\frac{2}{\pi ^{2}\beta ^{4}}\left[ -\frac{7}{%
720}\pi ^{4}+\frac{1}{24}\pi ^{2}\theta _{-}^{2}-\frac{1}{48}\theta
_{-}^{4}\right] +\frac{M^{2}}{2\pi ^{2}\beta ^{2}}\left[ \frac{1}{12}\pi
^{2}-\frac{1}{4}\theta _{-}^{2}\right] \nonumber\\
&&+\frac{1}{2\pi \beta ^{4}}\sum_{l}\,^{^{^{\prime }}}\left\{ \frac{1}{3}%
\left[ \left( \beta M\right) ^{2}+\left( \theta -\left( 2l-1\right) \pi
\right) ^{2}\right] ^{3/2}-\frac{1}{3}\left| \theta -\left( 2l-1\right) \pi
\right| ^{3}-\frac{1}{2}\left| \theta -\left( 2l-1\right) \pi \right| \beta
^{2}M^{2}-\frac{\left( \beta M\right) ^{4}}{16\pi \left| l\right| }\right\}
\nonumber\\
&&+\frac{M^{4}}{16\pi ^{2}}\left[ \ln \left( \frac{\beta M}{4\pi }\right)
+\gamma -\frac{3}{4}\right]
\end{eqnarray}
with $\theta _{-}$ now used. For fermions in the fundamental representation
of $SU(N)$, we may write 
\begin{eqnarray}
V_{FT} &=&-\frac{7\pi ^{2}N}{180\beta ^{4}}+\sum_{j}\frac{1}{12\pi ^{2}\beta
^{4}}\left[ 2\pi ^{2}\theta _{j}^{2}-\theta _{j}^{4}\right] +\frac{NM^{2}}{%
12\beta ^{2}}-\sum_{j}\frac{M^{2}}{4\pi ^{2}\beta ^{2}}
\theta _{j}^{2}\nonumber \\
&&+\frac{1}{\pi \beta ^{4}}\sum_{j,l}\,^{^{^{\prime }}}\left\{ \frac{1%
}{3}\left[ \left( \beta M\right) ^{2}+\left( \theta _{j}-\left( 2l-1\right)
\pi \right) ^{2}\right] ^{3/2}-\frac{1}{3}\left| \theta _{j}-\left(
2l-1\right) \pi \right| ^{3}-\frac{1}{2}\left| \theta _{j}-\left(
2l-1\right) \pi \right| \beta ^{2}M^{2}-\frac{\left( \beta M\right) ^{4}}{%
16\pi \left| l\right| }\right\} \nonumber\\
&&+\frac{NM^{4}}{8\pi ^{2}}\left[ \ln \left( \frac{\beta M}{4\pi }\right)
+\gamma -\frac{3}{4}\right] \text{.}
\end{eqnarray}
where the angles $\theta _{j}$ must now be chosen to lie between $-\pi $ and 
$\pi $.

\section{Conclusions}

We have found complete, simple expressions for $V_{B}\left(
\theta \right) $ and $V_{F}\left( \theta \right) $ 
in the high-temperature limit
which generalize
previously known expressions. Not only are the expressions simple, their
derivation is direct and relatively elementary. 
Our formulae reflect in a direct way periodicity in $ \theta$, a
property which is lost when analytically continuing power series
in $\beta \mu$ to $\beta \mu + i \theta$.

As a practical matter, it is natural to ask how accurate both the low- and
high-temperature expansions are. The low temperature expansion for $%
V_{B}\left( \theta \right) $ is an infinite series in $n$; using the first $%
10$ terms in the series gives an accuracy better than $1$ part in $10^{3}$
over the entire range $0$ to $2\pi $ for temperatures $T\leq 0.25M$. The
high temperature expansion also involves an infinite series, in the the
parameter $l$. In comparison, the high temperature expansion is within $5\%$
of the exact answer for all values of $\theta $ at $T=0.5M$ when terms up to 
$\left| l\right| =10$ are included. The accuracy improves substantially as $%
T\,\ $increases. Both expansions are more accurate when restricted to $%
\theta =0$.

Our primary interest in these results lies in their application
to the study of systems where a non-trivial Polyakov loop is
expected to occur.
The foremost physical example is QCD at finite temperature.
The high-temperature form of $V_B ( \theta )$ suggests that the
Bernoulli polynomials appear naturally in the free energy
of $SU(N)$ gauge theories with a non-trivial Polyakov loop,
essentially as
polynomials in the Polyakov loop eigenvalues.
In our recent work with Miller\cite{Meisinger:2001cq},
we have used this observation
to construct
a phenomenological free energy for the quark-gluon plasma
which reproduces much of the thermodynamic observed in
lattice simulations.
One can also apply the results obtained here 
to the Savvidy model 
at finite temperature
\cite{Starinets:1994vi,Meisinger:1997jt}.
Savvidy originally proposed a model of
the QCD vacuum in which gluons moved 
in a constant chromomagnetic field
\cite{Savvidy:1977as}.
Using low-temperature expansions, we have shown that a
confining state, where the Polyakov loop expectation value is zero,
minimizes the free energy. 
Using the Bessel function identities proven here,
we have recently
\cite{usrealsoonnow}  developed
a high-temperature expansion for this model which shows
that a non-zero Polyakov line is favored at high temperature.

\appendix
\section{Derivation of Low Temperature Expansions}

We begin by expanding the logarithms and performing the angular
integrations: 
\begin{eqnarray}
V_{B}\left( \theta \right)  &=&\frac{1}{\beta }\int \frac{d^{d}k}{\left(
2\pi \right) ^{d}}\ln \left[ 1-e^{-\beta \omega _{k}+i\theta }\right] +\frac{%
1}{\beta }\int \frac{d^{d}k}{\left( 2\pi \right) ^{d}}\ln \left[ 1-e^{-\beta
\omega _{k}-i\theta }\right]  \nonumber\\
&=&-\frac{4\pi ^{d/2}}{\Gamma (d/2)\left( 2\pi \right) ^{d}\beta }\int
dk\,k^{d-1}\sum_{n=1}^{\infty }\frac{1}{n}e^{-n\beta \omega _{k}}\cos \left(
n\theta \right) \text{.}
\end{eqnarray}
The standard substitution $k=M\sinh (t)$ gives 
\begin{eqnarray}
V_{B}\left( \theta \right)  &=&-\frac{4\pi ^{d/2}}{\Gamma (d/2)\left( 2\pi
\right) ^{d}\beta }\sum_{n=1}^{\infty }\frac{\cos \left( n\theta \right) }{n}%
M^{d}\int_{0}^{\infty }dt\,\cosh t\,\sinh t^{d-1}e^{-n\beta M\cosh t}
\nonumber \\
&=&\frac{4\pi ^{d/2}}{\Gamma (d/2)\left( 2\pi \right) ^{d}}%
\sum_{n=1}^{\infty }\frac{\cos \left( n\theta \right) }{n^{2}\beta ^{2}}M^{d}%
\frac{d}{dM}\int_{0}^{\infty }dt\,\,\sinh t^{d-1}e^{-n\beta M\cosh t} 
\nonumber\\
&=&\frac{4\pi ^{d/2}}{\Gamma (d/2)\left( 2\pi \right) ^{d}}%
\sum_{n=1}^{\infty }\frac{\cos \left( n\theta \right) }{n^{2}\beta ^{2}}M^{d}%
\frac{d}{dM}\left[ \frac{\Gamma \left( d/2\right) }{\sqrt{\pi }}\left( \frac{%
2}{n\beta M}\right) ^{\left( d-1\right) /2}K_{\left( d-1\right) /2}\left(
n\beta M\right) \right] 
\end{eqnarray}
This can in turn be reduced using standard recursion relations for modified
Bessel functions: 
\begin{eqnarray}
V_{B}\left( \theta \right)  &=&\frac{4\pi ^{\left( d-1\right) /2}}{\left(
2\pi \right) ^{d}\beta }M^{d}\sum_{n=1}^{\infty }\frac{\cos \left( n\theta
\right) }{n}\frac{d}{dz}\left[ \left( \frac{2}{z}\right) ^{\nu }K_{\nu
}\left( z\right) \right] _{z=n\beta M,\nu =\left( d-1\right) /2} \nonumber\\
&=&\frac{4\pi ^{\left( d-1\right) /2}}{\left( 2\pi \right) ^{d}\beta }%
M^{d}\sum_{n=1}^{\infty }\frac{\cos \left( n\theta \right) }{n}\left[ \left( 
\frac{2}{z}\right) ^{\nu }\left( \frac{dK_{\nu }\left( z\right) }{dz}-\frac{%
\nu }{z}K_{\nu }\left( z\right) \right) \right] _{z=n\beta M,\nu =\left(
d-1\right) /2} \nonumber\\
&=&\frac{4\pi ^{\left( d-1\right) /2}}{\left( 2\pi \right) ^{d}\beta }%
M^{d}\sum_{n=1}^{\infty }\frac{\cos \left( n\theta \right) }{n}\left[
-\left( \frac{2}{z}\right) ^{\nu }K_{\nu +1}\left( z\right) \right]
_{z=n\beta M,\nu =\left( d-1\right) /2} \nonumber\\
&=&-\frac{4\pi ^{\left( d-1\right) /2}}{\left( 2\pi \right) ^{d}\beta }%
M^{d}\sum_{n=1}^{\infty }\frac{\cos \left( n\theta \right) }{n}\left( \frac{2%
}{n\beta M}\right) ^{\left( d-1\right) /2}K_{\left( d+1\right) /2}\left(
n\beta M\right)  \nonumber\\
&=&-\frac{M^{d/2+1/2}}{2^{d/2-3/2}\pi ^{d/2+1/2}\beta ^{d/2+1/2}}%
\sum_{n=1}^{\infty }\frac{\cos \left( n\theta \right) }{n^{d/2+1/2}}%
K_{\left( d+1\right) /2}\left( n\beta M\right) 
\end{eqnarray}

\section{Proof of a Bessel Function Identity}

\vspace{1pt}In this appendix, we prove the Bessel function identity
\begin{equation}
\sum_{m=1}^{\infty }K_{0}(mr)\cos (m\phi )=\frac{1}{2}\left[ \ln \left( 
\frac{r}{4\pi }\right) +\gamma \right] +\frac{\pi }{2}\sum_{l}\,^{^{^{\prime
}}}\left[ \frac{1}{\sqrt{r^{2}+\left( \phi -2\pi l\right) ^{2}}}-\frac{1}{%
2\pi \left| l\right| }\right] \text{.}
\end{equation}
Using a standard integral representation \cite{GandR}
\begin{equation}
K_{0}(pz)=\int_{0}^{\infty }dt\,\frac{\cos (pt)}{\sqrt{t^{2}+z^{2}}}\text{,}
\end{equation}
we have 
\begin{eqnarray}
&&\sum_{m=1}^{\infty }\cos (m\phi )K_{0}(mr) \nonumber\\
&=&\sum_{m=1}^{\infty }\cos (m\phi )\int_{0}^{\infty }dk_{x}\frac{\cos
(k_{x}r)}{\sqrt{k_{x}^{2}+m^{2}}} \nonumber\\
&=&\frac{1}{2}\sum_{m=1}^{\infty }\cos (m\phi )\int dk_{x}\frac{1}{\sqrt{%
k_{x}^{2}+m^{2}}}e^{ik_{x}r} \nonumber\\
&=&\frac{1}{4\pi }\sum_{m\neq 0}\int dk_{x}dk_{y}\frac{1}{%
k_{x}^{2}+k_{y}^{2}+m^{2}}e^{ik_{x}r+im\phi }
\end{eqnarray}
We introduce a regulating mass $\mu $, which will be taken to zero at the
end of the calculation, obtaining 
\begin{equation}
\frac{1}{4\pi }\sum_{m\neq 0}\int dk_{x}dk_{y}\frac{1}{%
k_{x}^{2}+k_{y}^{2}+m^{2}+\mu ^{2}}e^{ik_{x}r+im\phi }
\end{equation}
We add and subtract the divergent $m=0$ term 
\begin{eqnarray}
&&\frac{1}{4\pi }\sum_{m}\int dk_{x}dk_{y}\frac{1}{%
k_{x}^{2}+k_{y}^{2}+m^{2}+\mu ^{2}}e^{ik_{x}r+im\phi }-\frac{1}{4\pi }\int
dk_{x}dk_{y}\frac{1}{k_{x}^{2}+k_{y}^{2}+\mu ^{2}}e^{ik_{x}r} \nonumber\\
&=&\frac{1}{4\pi }\sum_{m}\int dk_{x}dk_{y}dk_{z}\frac{1}{%
k_{x}^{2}+k_{y}^{2}+k_{z}^{2}+\mu ^{2}}\delta \left( k_{z}-m\right)
e^{ik_{x}r+ik_{z}\phi }-\frac{1}{4\pi }\int dk_{x}dk_{y}\frac{1}{%
k_{x}^{2}+k_{y}^{2}+\mu ^{2}}e^{ik_{x}r}
\end{eqnarray}
Using the Poisson summation technique in the form $\sum_{m}\delta \left(
k_{z}-m\right) =\sum_{n}\exp \left( -2\pi ink_{z}\right) $, we obtain 
\begin{equation}
\frac{1}{4\pi }\sum_{n}\int dk_{x}dk_{y}dk_{z}\frac{1}{%
k_{x}^{2}+k_{y}^{2}+k_{z}^{2}+\mu ^{2}}e^{ik_{x}r+ik_{z}\left( \phi -2\pi
n\right) }-\frac{1}{4\pi }\int dk_{x}dk_{y}\frac{1}{k_{x}^{2}+k_{y}^{2}+\mu
^{2}}e^{ik_{x}r}
\end{equation}
which reads in a compact notation 
\begin{equation}
\frac{\pi }{2}\sum_{n}\int \frac{d^{3}k}{\left( 2\pi \right) ^{3}}\frac{4\pi 
}{k^{2}+\mu ^{2}}e^{ik_{x}r+ik_{z}\left( \phi -2\pi n\right) }-\frac{1}{4\pi 
}\int d^{2}k\frac{1}{k^{2}+\mu ^{2}}e^{ik_{x}r}
\end{equation}
The first integral gives a sum of screened Coulomb, or Yukawa, potentials 
\begin{equation}
\frac{\pi }{2}\sum_{n}\left[ \frac{e^{-\mu \sqrt{r^{2}+\left( \phi -2\pi
n\right) ^{2}}}}{\sqrt{r^{2}+\left( \phi -2\pi n\right) ^{2}}}\right] -\frac{%
1}{4\pi }\int d^{2}k\frac{1}{k^{2}+\mu ^{2}}e^{ik_{x}r}
\end{equation}
and both terms appear to be problematic as $\mu \rightarrow 0$. The first
term can be made finite in this limit by subtracting the contribution at $%
r=\phi =0$ for $n\neq 0$. Using the notation $\sum_{n}{}^{^{^{\prime }}}$ to
denote a summation over all $n$ with the omission of the singular term when $%
n=0$, we have 
\begin{equation}
\frac{\pi }{2}\sum_{n}\,^{^{^{\prime }}}\left[ \frac{e^{-\mu \sqrt{%
r^{2}+\left( \phi -2\pi n\right) ^{2}}}}{\sqrt{r^{2}+\left( \phi -2\pi
n\right) ^{2}}}-\frac{e^{-\mu 2\pi \left| n\right| }}{2\pi \left| n\right| }%
\right] +\left[ \frac{\pi }{2}\sum_{n}\,^{^{^{\prime }}}\frac{e^{-\mu 2\pi
\left| n\right| }}{2\pi \left| n\right| }-\frac{1}{4\pi }\int d^{2}k\frac{1}{%
k^{2}+\mu ^{2}}e^{ik_{x}r}\right] \text{.}
\end{equation}
The second term in brackets can be evaluated by summing the series 
\begin{equation}
\frac{\pi }{2}\sum_{n}\,^{^{^{\prime }}}\frac{e^{-\mu 2\pi \left| n\right| }%
}{2\pi \left| n\right| }=-\frac{1}{2}\ln \left[ 1-e^{-2\pi \mu }\right] 
\end{equation}
and using the identity 
\begin{equation}
-\frac{1}{4\pi }\int d^{2}k\frac{1}{k^{2}+\mu ^{2}}e^{ik_{x}r}=-\frac{1}{2}%
K_{0}\left( \mu r\right) 
\end{equation}
so we have 
\begin{equation}
\frac{\pi }{2}\sum_{n}\,^{^{^{\prime }}}\left[ \frac{e^{-\mu \sqrt{%
r^{2}+\left( \phi -2\pi n\right) ^{2}}}}{\sqrt{r^{2}+\left( \phi -2\pi
n\right) ^{2}}}-\frac{e^{-\mu 2\pi \left| n\right| }}{2\pi \left| n\right| }%
\right] +\left[ -\frac{1}{2}\ln \left[ 1-e^{-2\pi \mu }\right] -\frac{1}{2}%
K_{0}\left( \mu r\right) \right] \text{.}
\end{equation}
In the limit $\mu \rightarrow 0$, the second term in brackets gives 
\begin{eqnarray}
&&-\frac{1}{2}\ln \left[ 1-e^{-2\pi \mu }\right] -\frac{1}{2}K_{0}\left( \mu
r\right)  \nonumber\\
&\rightarrow &-\frac{1}{2}\ln \left[ 2\pi \mu \right] +\frac{1}{2}\ln \left( 
\frac{\mu r}{2}\right) -\frac{1}{2}\psi (1)=\frac{1}{2}\ln \left( \frac{r}{%
4\pi }\right) +\frac{1}{2}\gamma 
\end{eqnarray}
so we finally obtain 
\begin{equation}
\sum_{m=1}^{\infty }K_{0}(mr)\cos (m\phi )=\frac{\pi }{2}\sum_{n}\,^{^{^{%
\prime }}}\left[ \frac{1}{\sqrt{r^{2}+\left( \phi -2\pi n\right) ^{2}}}-%
\frac{1}{2\pi \left| n\right| }\right] +\frac{1}{2}\ln \left( \frac{r}{4\pi }%
\right) +\frac{1}{2}\gamma \text{.}
\end{equation}

\end{document}